\documentclass[aps,prd,draft,amsmath,amssymb,preprintnumbers,nofootinbib,a4paper,11pt]{revtex4}
\pdfoutput=1
\usepackage{amsfonts,amsmath,amssymb,mathrsfs}
\usepackage{hyperref}
\usepackage{subfigure}
\usepackage{color}
\usepackage{graphicx}  
\usepackage{dcolumn}   
\usepackage{bm}        
\usepackage[english]{babel}

\begin{document}

\title{\Large{Inflating with the Composite Inflaton}}

\author{Phongpichit Channuie}\email{channuie@cp3.dias.sdu.dk}
\affiliation{{\rm CP}$^{\bf 3}${\rm-Origins} \& the Danish Institute for Advanced Study {\rm DIAS},\\ 
University of Southern Denmark, Campusvej 55, \\DK-5230 Odense M, Denmark}

\date{\today}

\begin{abstract}

The idea of the strongly interacting field theories constitutes one of the pillars of the standard model of particle physics. The goal of this work is to export such theories to cosmology. Here we are interested in models in which the inflaton is a composite object, instead of a fundamental one. In so doing, we first construct the general setup, in the metric formulation, for generic models of composite inflation. We subsequently introduce the relevant examples in which the inflaton is identified with: A) the lightest (composite) scalar field in the minimal walking technicolor theory, and B) the glueball field of a pure Yang-Mills theory. We demonstrate that they were both viable to achieve successful inflation. Surprisingly, we find that the scale of composite inflation is of the order of the expected grand unification scale, $10^{16}$ GeV. \\ 
[.1cm]
{\footnotesize  \it Preprint: CP$^3$-Origins-2012-004 \& DIAS-2012-5}

\end{abstract}

\maketitle

\section{Introduction}
\label{intro}

The origin of the Universe is one of the ultimate questions in Science. Our understanding is still obscured by surprisingly many layers of its mystery. According to the cosmological viewpoint, the Big Bang model is extremely successful in explaining a remarkably broad range of observations. Although it has never claimed to elucidate the final truth, it provides a framework for understanding the evolution of the Universe from the few fractions of a second of its existence till the present. Its predictions also are, for instance, the abundances of the primordially synthesised light elements; the thermal relic of the Big Bang in the form of an isotropic microwave background having a blackbody-like spectrum; and the matter structures formed by gravitational collapse from primordial fluctuations.

Nevertheless, the theory left some unanswered puzzles concerning its initial conditions. These, for example, include the origin of the homogeneity and flatness of spatial sections; the origin of matter and radiation; the origin of matter-antimatter asymmetry; the origin of the primordial seeds for structure; the origin and nature of the dark matter and dark energy.

Traditionally, the initial conditions are well described within a framework of inflationary cosmology. The inflationary paradigm exported new ideas from particle physics to theoretical cosmology, e.g.\,\cite{Starobinsky:1979ty,Starobinsky:1980te,Mukhanov:1981xt,Guth:1980zm,Linde:1981mu,Albrecht:1982wi}. Similar to the standard model Higgs mechanism, inflation is also modeled by introducing, at least, a new scalar field dubbed the \lq\lq inflaton\rq\rq. However, a (fundamental) scalar field sector in field theories is plagued by the so-called Òhierarchy problemÓ. Commonly, this means that quantum corrections generate unprotected quadratic divergences which must be fine-tunned away. Similarly, the inflaton also suffers from the same kind of untamed quantum corrections.

Having shown that the models of composite inflation, in the metric formulation, lead to a consistent picture  \cite{Bezrukov2011}, here we will use such a formulation to demonstrate that the models below are both possible to achieve successful inflation.

In this work, we investigate a number of models of inflation in which the inflaton is designed to be the lightest composite state present in different kinds of four-dimensional strongly coupled theories.  We will start here by briefly reviewing the general setup for generic models of composite inflation \cite{Bezrukov2011}. We then introduce the relevant examples in which the inflaton is identified with: A) the lightest (composite) scalar field in the minimal walking technicolor theory \cite{Channuie:2011rq}, and B) the glueball field of a pure Yang-Mills theory \cite{Bezrukov2011}. Finally, we briefly summarize our finding in the last section.

\section{Composite Inflation (CI) Setup}
\label{setup}

For a generic strongly coupled theory, we start by identifying the inflaton with one of the lightest composite states of the theory, $\Phi$. This state has mass dimension $d$. This is the physical dimension coming from the sum of the engineering dimensions of the elementary fields constituting the inflaton, augmented by the anomalous dimensions due to quantum corrections in the underlying gauge theory. In the metric formulation, we consider the following coupling to gravity in the Jordan (J) frame \cite{Bezrukov2011}: 
\begin{equation}
\mathcal{S}^{\rm J}_{\rm CI}=\int d^{4}x \sqrt{-g}\left[-\frac{M^{2}+\xi\,{\Phi}^{\frac{2}{d}}}{2}g^{\mu\nu}R_{\mu\nu}+\mathcal{L}_{\Phi}\right],\,\,\,\mathcal{L}_{\Phi}=g^{\mu\nu}\Phi^{\frac{2-2d}{d}}\partial_{\mu}\Phi\partial_{\nu}\Phi-\mathcal{V}({\Phi}), \label{nonminimal}
\end{equation}
with $\mathcal{L}_{\Phi}$ the low energy effective Lagrangian for the field $\Phi$ constrained by the symmetries of the underlying strongly coupled theory.  In this investigation, $M$ is not automatically the Planck constant $M_{\rm Pl}$. The non-minimal coupling to gravity is parametrized by the dimensionless coupling $\xi$. The non-analytic power of $\Phi$ may emerge because we are requiring a dimensionless coupling with the Ricci scalar.

With the knowledge of the conformal transformation:
\begin{equation}
g_{\mu\nu}\rightarrow\tilde{g}_{\mu\nu}=\Omega({\Phi})^2 g_{\mu\nu},\quad\Omega({\Phi})^2=\frac{M^2+\xi\Phi^{\frac{2}{d}}}{M_{\rm Pl}^2},
\end{equation}
we can diagonalize the gravity-composite dynamics model such that 
\begin{equation}
\quad\tilde{g}^{\mu \nu}=\Omega(\Phi)^{-2}g^{\mu\nu},\quad\sqrt{-\tilde{g}}=\Omega(\Phi)^4\sqrt{-g}.
\end{equation} 
 
After imposing the conformal transformation, we come up with the action in the Einstein (E) frame:
\begin{equation}
\mathcal{S}^{\rm E}_{\rm CI} =\int d^{4}x \sqrt{-g}\left[ -\frac{1}{2} M_{\rm Pl}^2 \,\, \tilde{g}^{\mu \nu}\tilde{R}_{\mu \nu} + \Omega^{-2} \left(\Phi^{\frac{2-2d}{d}} + 3 M^{2}_{\rm Pl} {\Omega'}^{2} \right)\tilde{g}^{\mu \nu} \partial_{\mu} \Phi \partial_{\nu} \Phi  - \Omega ^{-4} \mathcal{V}({\Phi}) \right].
\end{equation}

Primes denotes derivatives with respect to $\Phi$, and tildes represent the parameters in the Einstein frame. To deal with an involved kinetic term for the inflaton, it is convenient to introduce a canonically normalized field $\chi$ related to $\Phi$ via
\begin{equation}
\frac{1}{2} \tilde{g}^{\mu \nu} \partial_{\mu} \chi (\Phi) \partial_{\nu} \chi(\Phi) = \frac{1}{2} \left( \frac{d \chi}{d \Phi} \right)^2 \tilde{g}^{\mu \nu} \partial_{\mu} \Phi \partial_{\nu} \Phi \ ,
\end{equation}
with
\begin{equation}
\frac{1}{2} \left( \frac{d \chi}{d \Phi} \right)^2 = \Omega^{-2}\left(1 +  \frac{3 \xi ^2}{d^2 M_{\rm Pl}^2} \Omega^{-2} \Phi^{\frac{2}{d}} \right) \Phi ^{\frac{2-2d}{d}}. \label{defchi}
\end{equation}

In terms of the canonically normalized field, we have: 
\begin{equation}
\mathcal{S}^{\rm E}_{\rm CI} =\int d^{4}x \sqrt{-g}\left[-\frac{1}{2} M_{\rm Pl}^2 R+ \frac{1}{2} g^{\mu \nu} \partial_{\mu} \chi \partial_{\nu} \chi- U(\chi)  \right].
\end{equation}

With
\begin{equation}
U(\chi) \equiv \Omega^{-4}\mathcal{V}(\Phi(\chi)).  
\end{equation}

Now we have dropped tildes for later convenience. We will analyze the dynamics in the Einstein frame, and therefore define the slow-roll parameters in terms of $U$ and $\chi$:
\begin{equation}
\epsilon = \frac{M_{\rm Pl}^2}{2} \left( \frac{dU / d \chi}{U} \right)^2, \,\,\eta = M_{\rm Pl}^2 \left( \frac{d^2U / d \chi^2}{U} \right), \,\,N = \frac{1}{M_{\rm Pl}^2} \int _{\chi_{end}} ^{\chi_{ini}} \frac{U}{dU /d\chi} d \chi. \label{epsilon}
\end{equation}
 
However, the above parameters can be cast in terms of $\Phi$ such that we do not need an explicit solution of (\ref{defchi}). We simply obtain:
\begin{equation}
\epsilon = \frac{1}{4} \frac{\left( \left( 1+ \frac{\xi}{M^{2}} \Phi^{\frac{2}{d}}\right)\Phi \frac{\mathcal{V}'}{\mathcal{V}}- \frac{4}{d} \frac{\xi}{M^{2}} \Phi^{\frac{2}{d}} \right)^2}{\left(1 + \frac{\xi}{M^{2}} \Phi^{\frac{2}{d}} \right)\frac{1}{M^{2}}\Phi^{\frac{2}{d}} + \frac{3}{d^2}\left( \frac{\xi}{M^{2}} \Phi^{\frac{2}{d}}\right)^2}\ , \label{setep}
\end{equation}
\begin{align}
\eta&=M_{\rm Pl}^2\left\{\frac{\mathcal{V}''}{\mathcal{V}}-4\Omega^{-1}\left(\Omega^{-1}\frac{\xi}{dM^{2}_{\rm Pl}}\left(\frac{2-d}{d}\right)\Phi^{\frac{2-2d}{d}}-\Omega^{-3}\left(\frac{\xi}{dM^{2}_{\rm Pl}}\right)^{2}\Phi^{\frac{4-2d}{d}}\right)\right\}\left(\frac{d\Phi}{d\chi}\right)^{2}\nonumber\\&-M_{\rm Pl}^2\left[\left\{8\Omega^{-1}\left(\frac{1}{\Omega}\frac{\xi\Phi^{\frac{2-d}{d}}}{dM^{2}_{\rm Pl}}\right)\frac{\mathcal{V}'}{\mathcal{V}}-20\Omega^{-2}\left(\frac{1}{\Omega}\frac{\xi\Phi^{\frac{2-d}{d}}}{dM^{2}_{\rm Pl}}\right)^{2}\right\}\left(\frac{d\Phi}{d\chi}\right)^{2}+\left\{\frac{\mathcal{V}'}{\mathcal{V}}-4\Omega^{-2}\frac{\xi\Phi^{\frac{2-d}{d}}}{dM^{2}_{\rm Pl}}\right\}\frac{d^{2}\Phi}{d\chi^{2}}\right] \ , \label{STeta}
\end{align}
\begin{equation}
N = \frac{2}{M^{2}}\int_{\Phi_{end}}^{\Phi_{ini}} \frac{\Phi^{\frac{2-d}{d}}\left(1+\frac{3 \xi^2}{d^2M^{2}} \Phi^{\frac{2}{d}} \frac{1}{1+ \frac{\xi}{M^{2}}\Phi^{\frac{2}{d}}}\right)}{\frac{-4}{d}\frac{\xi}{M^{2}}\Phi^{\frac{2}{d}}+ \left(1+ \frac{\xi}{M^{2}}\Phi^{\frac{2}{d}} \right)\Phi\frac{\mathcal{V}'}{\mathcal{V}}} d \Phi \ . \label{efold}
\end{equation}

Here we spelled out the setup for generic models of composite inflation. We will use our setup for investigating different underlying models of composite inflation.

\section{Composite Models of Inflation}
\label{model}

In this work, our inflaton stems from a natural four-dimensional dynamics, and therefore it is free from unnatural fine-tuning. Having written the generic form of the relevant parameters for inflation, we now introduce some relevant examples.\\

\subsection{Minimal Composite Inflation (MCI)}
\label{model1}

Begging the Lagrangian from Minimal Walking Technicolor \cite{Foadi:2007ue}, the Higgs Lagrangian is now identified with the MCI effective theory in which we couple non-minimally to gravity in the Jordan frame as follows:
\begin{equation}
\mathcal{S}^{\rm J}_{\rm MCI}=\int d^{4}x \sqrt{-g}\left[- \frac{M^{2}_{\rm Pl}}{2}R - \frac{\xi}{2}\text{Tr}\left(\mathcal{M}\mathcal{M}^{\dagger}\right)R+\mathcal{L}_{\text{MCI}}\right]. \label{tcsj}
\end{equation}

Here the MCI Lagrangian $\mathcal{L}_{\rm MCI}$ is given by
\begin{equation}
\mathcal{L}_{\text{MCI}}=\frac{1}{2}{\rm Tr}\left[D_{\mu}\mathcal{M}D^{\mu}\mathcal{M}^{\dagger}\right]-\mathcal{V}(\mathcal{M}), \label{tcsj1}
\end{equation}
where the potential reads
\begin{align}
\mathcal{V}(\mathcal{M})=&-\frac{m^{2}}{2}{\rm Tr}\left(\mathcal{M}\mathcal{M}^{\dagger}\right)+\frac{\lambda}{4}{\rm Tr}\left(\mathcal{M}\mathcal{M}^{\dagger}\right)^{2}+\lambda'{\rm Tr}\left(\mathcal{M}\mathcal{M}^{\dagger}\mathcal{M}\mathcal{M}^{\dagger}\right)\nonumber\\&-2\lambda''\left[{\rm Det}(\mathcal{M})+{\rm Det}(\mathcal{M}^{\dagger})\right], \label{tcsj2}
\end{align}
with 
\begin{equation}
\mathcal{M}=\left[\frac{\sigma+i\Theta}{2}+\sqrt{2}\left(i\Pi^{a}+\tilde{\Pi}^{a}\right)X^{a}\right]\mathcal{E}\,\,,\,a=1...9\,. \label{toto}
\end{equation}

The $X^{a}$'s are the generators of the $SU(4)$ group which do not leave the vacuum expectation value (vev) of $\mathcal{M}$ invariant
\begin{equation}
\left<\mathcal{M}\right>=\frac{v}{2}\mathcal{E}\,. \label{toto11}
\end{equation}

The matrix $\mathcal{E}$ is a $4\times 4$ matrix defined in terms of the 2-dimensional unit matrix as
\begin{equation}
\mathcal{E}=
\begin{pmatrix}0 & 1 \\1 & 0 \end{pmatrix}. \label{tot2o}
\end{equation}

Here the inflaton is assigned to be the field $\sigma$. The other scalars are the nine goldstone bosons, ($\Pi^a$), in which we assume to become the longitudinal degrees of freedom of the conveniently gauged $SU(4)$  flavor symmetry.  When the techni-fermionic condensate is dynamically generated, the $SU(4)$ gauge symmetry spontaneously breaks to $SO(4)$. The remaining composite scalars $\Theta$ and $\widetilde{\Pi}^{a}$ are massive, and for (near) conformal field theories, expected to be heavier than $\sigma$. Therefore it is sensible to consider the $\sigma$  dynamics first.

Now we drop the other fields in $\mathcal{M}$. In the general framework outlined above, this field, $\Phi\equiv\sigma$, has mass dimension $d = 1$. So the relevant composite inflaton effective action reads: 
\begin{equation}
\mathcal{S}^{\rm J}_{\rm MCI}=\int d^{4}x \sqrt{-g}\left[-\frac{M^{2}+\xi\sigma^{2}}{2}R + \frac{1}{2}g^{\mu\nu}\partial_{\mu}\sigma\partial_{\nu}\sigma-\mathcal{V}_{\rm MCI}(\sigma)\right], \label{sjtc}
\end{equation}
where
\begin{equation}
\mathcal{V}_{\rm MCI}(\sigma)=-\frac{m^{2}}{2}\sigma^{2}+\frac{\kappa}{4}\sigma^{4}. \label{kap} 
\end{equation}

Here the linear combination $\kappa=\left(\lambda+\lambda^{'}-\lambda^{''}\right)$ is the composite inflaton self-coupling. Imposing the conformal transformation $g_{\mu \nu} \rightarrow \tilde{g}_{\mu \nu} = \Omega ^2 g_{\mu \nu}$, we can eliminate the non-minimal coupling between $\sigma$ and the gravitational field. The resulting action in the Einstein frame is:
 \begin{equation}
\mathcal{S}^{\rm E}_{\rm MCI}=\int d^{4}x \sqrt{-g}\left[- \frac{M^{2}_{\rm Pl}}{2}R+\Omega^{-2}\left(1+\frac{3\Omega^{-2}\xi^{2}\sigma^{2}}{M^{2}_{\rm Pl}}\right)g^{\mu\nu}\partial_{\mu}\sigma\partial_{\nu}\sigma-\Omega^{-4}\mathcal{V}_{\rm MCI}(\sigma)\right] , \label{action}
\end{equation}
where
\begin{equation}
\Omega^{2}=\frac{M^{2}+\xi\sigma^{2}}{M^{2}_{\rm Pl}}. \label{actiont}
\end{equation}

We are now able to determine the slow-roll parameters and the constraints relevant for inflation. Now we consider the large field approximation, i.e.,
\begin{equation}
 \sigma \gg \frac{M}{\sqrt{\xi}} \ . 
 \label{lfa}
 \end{equation}

From (\ref{setep}), we obtain
\begin{equation}
\epsilon\simeq \frac{4\,M^{4}}{\xi^{2}\left(\xi^{-1}+3\right)\sigma^{4}}. \label{epsi-app}
\end{equation}

Inflation ends when $\epsilon=1$, so that:
\begin{equation}
\sigma_{\rm end}\simeq\frac{M}{\sqrt{\xi}}\left(\frac{4}{\left(\xi^{-1}+3\right)}\right)^{\frac{1}{4}}. \label{epsi-end}
\end{equation}

In the large field limit, the number of e-foldings (\ref{efold}) is:
\begin{equation}
\mathcal{N}  
\simeq \frac{\xi\left(\xi^{-1}+3\right)}{4\,M^{2}}\left(\sigma^{2}_{\rm ini}-\sigma^{2}_{\rm end}\right). \label{N-MCI}
\end{equation}

Now we deduce that $\sigma^{2}_{\rm ini}\gg\sigma^{2}_{\rm end}$. So we obtain
\begin{equation}
\sigma^{2}_{\rm ini}\simeq\frac{4\,M^{2}\,\mathcal{N}}{\xi\left(\xi^{-1}+3\right)}. \label{ini-MCI}
\end{equation}

Notice that we recover the results presented in \cite{Channuie:2011rq} by imposing $M=M_{\rm Pl}$, and using $\xi\gg 3$ in (\ref{epsi-end}) and (\ref{ini-MCI}). To generate the proper amplitude of the density perturbations, the potential must satisfy at $\sigma_{\rm WMAP}$ the normalization condition \cite{arXiv:0812.3622}: 
\begin{equation}
\frac{U_{\rm ini}}{\epsilon_{\rm ini}} \simeq (0.0276 M_{\rm Pl})^4 \ , 
\end{equation}
corresponding to the initial value assumed by the inflaton. In the large field limit, we find
\begin{equation}
U_{\rm ini}  
 \simeq \frac{\kappa M^{4}_{\rm Pl}}{4\xi^{2}} \ .
\end{equation} 
while:
\begin{equation}
\epsilon_{\rm ini} \simeq \frac{4\,M^{4}}{\xi^{2}\left(\xi^{-1}+3\right)\sigma^{4}_{\rm ini}}\ .
\end{equation}

Using $\mathcal{N}\approx60$, we can now determine the magnitude of the non-minimal coupling: 
\begin{equation}
\xi = \frac{\mathcal{N}}{(0.0276)^2}\sqrt{\frac{\kappa}{3}}  \sim 46000 \sqrt{\kappa}\ .
\end{equation}

For a strongly coupled theory we expect $\kappa$ to be of the order of unity and therefore $\xi \sim 46000$. This analysis resembles very closely the one for the SM Higgs inflation \cite{Bezrukov:2007ep}, except that our effective theory for the composite inflaton cannot be used for arbitrary large value of scalar field. The effective theory  is valid for: 
\begin{equation}
 \sigma < 4\pi v \  ,
\end{equation}
implying
\begin{equation}
v >  \frac{\sqrt{80}\,M}{4\pi \sqrt{\xi}} \sim 0.81 \times 10^{16} ~{\rm  GeV}\,.
\end{equation}
with the above value obtained for the reduced Planck mass $M_{\rm Pl}$ of $M\equiv M_{\rm Pl}=2.44\times 10^{18}$~GeV. This value is surprisingly close to the typical grand unification scale, $10^{16}$\, GeV. We noticed that this phenomenological constraint on $v$ forbids the identification of the composite inflaton with  the composite Higgs.

\subsection{Glueball Inflation (GI)}
\label{model2}

Pure Yang-Mills theories featuring only gluonic-type fields are the simplest examples of strongly coupled theories. In this case, the candidate for the inflaton is the interpolating field describing the lightest glueball. 
\begin{equation}
\Phi\equiv\varphi(x) = \frac{\beta}{g} {\rm Tr} \left[ G^{\mu \nu} G_{\mu \nu}\right] \ ,
\end{equation}
where $G^{\mu \nu}$ is the standard non-abelian field strength and $\beta$ is the full beta function of the theory in any renormalization scheme.  The Yang-Mills trace anomaly constrains the low energy effective Lagrangian for the lightest glueball state \cite{Schechter:1980ak,Migdal:1982jp,Cornwall:1983zb} to be: 
\begin{equation}
\mathcal{L}_{\rm GI} = \varphi^{-\frac{3}{2}}\partial_{\mu}\varphi \partial^{\mu} \varphi  - \mathcal{V}_{ \rm GI} (\varphi), \quad \mathcal{V}_{\rm GI} (\varphi) =  \frac{\varphi}{2} \ln \left( \frac{\varphi}{\Lambda^4}\right)  \ .
\end{equation}

In the general framework outlined above, this field has mass dimension $d=4$, and the non-minimally coupled glueball effective action to gravity reads: 
\begin{equation}
\mathcal{S}^{\rm J}_{\rm GI}=\int d^{4}x \sqrt{-g}\left[-\frac{M^{2} +  \xi  \varphi^{\frac{1}{2}}}{2}R + \varphi ^{-\frac{3}{2}} \partial_{\mu} \varphi \partial^{\mu} \varphi - \mathcal{V}_{ \rm GI} (\varphi) \right]. \label{gLH}
\end{equation}

Imposing the conformal transformation with 
\begin{equation}
\Omega^2 = \frac{M^{2} + \xi \varphi ^{\frac{1}{2}}}{M_p ^2}  \ ,
\end{equation}
 the action in the Einstein frame becomes: 
\begin{equation}
\mathcal{S}^{\rm E}_{\rm GI}=\int d^{4}x \sqrt{-{g}}\left[-\frac{M_{\rm Pl}^{2}}{2}R+\Omega^{-2}\left(1+\frac{3\Omega^{-2}\xi^2 \varphi^{\frac{1}{2}}}{16M^{2}_{\rm Pl}}\right)g^{\mu\nu}\,\, \varphi^{-\frac{3}{2}}\partial_{\mu}\varphi \partial^{\nu}\varphi-\Omega^{-4}\mathcal{V}_{\rm GI} (\varphi)\right]. \label{Gnonminimal}
\end{equation}

We are now able to determine the slow-roll parameters and constraints relevant for inflation. Here we consider the large field regime, i.e.: 
 \begin{equation}
 \varphi^{\frac{1}{2}} \gg \frac{M^{2}}{\xi} \ . 
 \label{lfe}
 \end{equation}
 
In this limit, from (\ref{setep}), we obtain::
\begin{equation}
\epsilon \simeq 
 \frac{1}{ 4 \ln \left( \frac{\varphi}{\Lambda ^4} \right)^2 \left( \xi^{-1} + \frac{3}{16} \right)}.
\end{equation}

Inflation ends when $\epsilon =1$  such that:
\begin{equation}
 \frac{\varphi_{\rm end}}{\Lambda^4} = \exp \left( \frac{1}{2 \sqrt{\left(\xi ^{-1}  + \frac{3}{16}\right)}} \right) \label{phiend}.
\end{equation}

In the large field limit, the number of e-foldings (\ref{efold}) is:
\begin{equation}
\mathcal{N}  
\simeq \left[  \left(\xi^{-1}+\frac{3 }{16} \right) \ln \left( \frac{\varphi}{\Lambda^4}\right)^2 \right]_{\varphi_{\rm end}}^{\varphi_{\rm ini}}.
\end{equation}

A simple way to determine the value of $\varphi_{\rm ini}$ associated to the moment in whic inflation starts is to require a minimal numbers of e-foldings compatible with a successful inflation, i.e. $\mathcal{N}\approx60$. This leads to: 
\begin{equation} 
\frac{\varphi_{\rm ini}}{\Lambda^4}  \simeq \exp \left( \sqrt{\frac{60}{\xi^{-1}+\frac{3}{16}} }\right) \ .
\end{equation}

Further relevant informations can be extracted using the  WMAP \cite{arXiv:0812.3622} normalization condition:
\begin{equation}
\frac{U_{\rm ini}}{\epsilon_{\rm ini} } = (0.0276 M_{\rm Pl})^4.
\end{equation}

The label \lq\lq ini\rq\rq signifies that this expression has to be evaluated at the beginning of the inflationary period. This condition helps us estimating the magnitude of the non-minimal coupling. We deduce: 
\begin{equation}
U_{\rm ini}  
 \simeq \frac{M_{\rm Pl}^4}{2 \xi^2} \ln \left( \frac{\varphi_{\rm ini}}{\Lambda^4} \right)  
\simeq \frac{M_{\rm Pl}^4}{2 \xi^2} \sqrt{\frac{60}{\xi^{-1}+ 0.1875}}\,,
\end{equation} 
while:
\begin{equation}
\epsilon_{\rm ini} \simeq \frac{1}{ 4 \ln \left( \frac{\varphi_{\rm ini}}{\Lambda ^4} \right)^2 \left( \xi^{-1} + \frac{3}{16} \right)} = 0.0041\ .
\end{equation}

We can therefore determine the magnitude of the non-minimal coupling which assumes the following value:  
\begin{equation}
\xi \simeq 6.1 \times 10^{4}.
\end{equation}

The knowledge of the non-minimal coupling allows us to estimate the initial and final value of the composite glueball field $\varphi$. We have in units of the strong scale $\Lambda$: 
\begin{equation}
\frac{\varphi_{\rm end}^{\frac{1}{4}}}{\Lambda} \sim 1.3\,\, ,\qquad \,\,\,\,\, \frac{\varphi_{\rm ini}^{\frac{1}{4}}}{\Lambda}\sim 88\,.
\end{equation}

To be more concrete, let us further relate the strongly coupled scale $\Lambda$ with $M$ recalling that we are working in the large field regime (\ref{lfe}). This implies that the  smallest value assumed by the inflaton must satisfy (\ref{lfe}) and therefore we obtain: 
\begin{equation}
\Lambda >\frac{M}{\sqrt{\xi}} \ .
\end{equation}
Here $M$ is the reduced Planck mass $2.44\times 10^{18}$ GeV yielding: 
\begin{equation}
 \Lambda > 0.9 \times 10^{16}\,\,{\rm GeV} \ .
\end{equation}
This is the typical scale for grand unification, in complete agreement with our earlier results for the first  model of composite inflation explored in the previous section for a very different underlying model of composite inflation.

\section{Conclusion}
\label{conclude}

We constructed the general setup, in the metric formulation, for generic models of composite inflation. We then introduced models in which the inflatons are the composite fields stemming from four dimensional strongly interacting theories. We demonstrated that they are both viable to achieve successful inflation. Surprisingly, we discovered that the scale of composite inflation is of the order of the expected grand unification scale, $10^{16}$\,GeV.

\section{Acknowledgments}

The work is in collaboration with J. J. J\"{o}rgensen and F. Sannino. It is a pleasure to thank J. Channuie, E. D. Nobile, N. Kaewkao, F. Sannino, A. Shunava and J. Virkaj\"{a}rvi for comments and careful reading the manuscript. PC is fully granted by the Royal Thai Government under the program {\em Strategic Scholarships for Frontier Research Network} of the Thailand's  Commission for Higher Education.


\begin{thebibliography}{100}  

\bibitem{Starobinsky:1979ty}
 A.~A.~Starobinsky, ``Relict Gravitation Radiation Spectrum and Initial State of the Universe. (In Russian),''
 JETP Lett.\  {\bf 30}, 682 (1979)
 [Pisma Zh.\ Eksp.\ Teor.\ Fiz.\  {\bf 30}, 719 (1979)].

\bibitem{Starobinsky:1980te}
 A.~A.~Starobinsky, ``A New Type of Isotropic Cosmological Models Without Singularity,''
 Phys.\ Lett.\  B {\bf 91}, 99 (1980).

\bibitem{Mukhanov:1981xt}
 V.~F.~Mukhanov and G.~V.~Chibisov, ``Quantum Fluctuation and Nonsingular Universe. (In Russian),''
 JETP Lett.\  {\bf 33}, 532 (1981)
 [Pisma Zh.\ Eksp.\ Teor.\ Fiz.\  {\bf 33}, 549 (1981)].

\bibitem{Guth:1980zm}
 A.~H.~Guth, ``The Inflationary Universe: A Possible Solution to the Horizon and Flatness Problems,''
 Phys.\ Rev.\  D {\bf 23}, 347 (1981).

\bibitem{Linde:1981mu}
 A.~D.~Linde, ``A New Inflationary Universe Scenario: A Possible Solution of the Horizon, Flatness, Homogeneity, Isotropy and Primordial Monopole Problems,''
 Phys.\ Lett.\  B {\bf 108}, 389 (1982).

\bibitem{Albrecht:1982wi}
 A.~Albrecht and P.~J.~Steinhardt, ``Cosmology for Grand Unified Theories with Radiatively Induced Symmetry Breaking,''
 Phys.\ Rev.\ Lett.\  {\bf 48}, 1220 (1982).

\bibitem{Channuie:2011rq}
 P.~Channuie, J.~J.~Joergensen and F.~Sannino,
 ``Minimal Composite Inflation,''
 JCAP {\bf 1105}, 007 (2011)
 [arXiv:1102.2898 [hep-ph]].

\bibitem{Bezrukov2011}
  F.~Bezrukov, P.~Channuie, J.~J.~Joergensen and F.~Sannino, ``Composite Inflation Setup and Glueball Inflation,''
  arXiv:1112.4054 [hep-ph].
 
\bibitem{arXiv:0812.3622}
 F.~Bezrukov, D.~Gorbunov and M.~Shaposhnikov,
 ``On initial conditions for the Hot Big Bang,''
 JCAP {\bf 0906}, 029 (2009)
 [arXiv:0812.3622 [hep-ph]].
 
 \bibitem{Schechter:1980ak}
 J.~Schechter,
 ``Effective Lagrangian with Two Color Singlet Gluon Fields,''
 Phys.\ Rev.\  D {\bf 21}, 3393 (1980).

\bibitem{Migdal:1982jp}
 A.~A.~Migdal and M.~A.~Shifman,
 ``Dilaton Effective Lagrangian in Gluodynamics,''
 Phys.\ Lett.\  B {\bf 114}, 445 (1982).

\bibitem{Cornwall:1983zb}
 J.~M.~Cornwall and A.~Soni,
 ``Couplings of Low Lying Glueballs to Light Quarks, Gluons and Hadrons,''
 Phys.\ Rev.\  D {\bf 29}, 1424 (1984). 
 
\bibitem{Foadi:2007ue} 
  R.~Foadi, M.~T.~Frandsen, T.~A.~Ryttov and F.~Sannino,``Minimal Walking Technicolor: Set Up for Collider Physics,''
  Phys.\ Rev.\ D {\bf 76}, 055005 (2007)
  [arXiv:0706.1696 [hep-ph]].
  
\bibitem{Bezrukov:2007ep} 
  F.~L.~Bezrukov and M.~Shaposhnikov, ``The Standard Model Higgs boson as the inflaton,''
  Phys.\ Lett.\ B {\bf 659}, 703 (2008)
  [arXiv:0710.3755 [hep-th]].

\end{thebibliography}
\end{document}